\documentstyle{lamuphys}

\begin{document}

\title{Localized Flux Lines and the Bose Glass}

\author{Uwe C. T\"auber}

\institute{University of Oxford, 
	Department of Physics -- Theoretical Physics, 
	1 Keble Road, Oxford OX1 3NP, U.K.}


\maketitle

\begin{abstract}

Columnar defects provide effective pinning centers for magnetic flux
lines in high--$T_{\rm c}$ superconductors.
Utilizing a mapping of the statistical mechanics of directed lines to
the quantum mechanics of two--dimensional bosons, one expects an
entangled flux liquid phase at high temperatures, separated by a
second--order localization transition from a low--temperature ``Bose
glass'' phase with infinite tilt modulus.
Recent decoration experiments have demonstrated that below the
matching field the repulsive forces between the vortices may be
sufficiently large to produce strong spatial correlations in the Bose
glass.
This is confirmed by numerical simulations, and a remarkably wide soft
``Coulomb gap'' at the chemical potential is found in the distribution
of pinning energies.
At low currents, the dominant transport mechanism in the Bose glass
phase proceeds via the formation of double kinks between not
necessarily adjacent columnar pins, similar to variable--range hopping
in disordered semiconductors.
The strong correlation effects originating in the long--range vortex
interactions drastically reduce variable--range hopping transport.

\end{abstract}

\section{Pinning of Flux Lines to Columnar Defects}

The remarkably rich phase diagram of magnetic flux lines in
high--$T_{\rm c}$ superconductors, especially when subject to point
and / or extended disorder, has attracted considerable experimental
and theoretical interest (for a recent review, see
Ref.~\cite{review}). 
For the purpose of applying type--II superconductors in external
magnetic fields, an effective vortex pinning mechanism is essential,
in order to minimize dissipative losses caused by the Lorentz--force
induced movement of flux lines across the sample. 
This issue becomes even more important for the high--$T_{\rm c}$
superconductors, because the strongly enhanced thermal fluctuations in
these materials render the Abrikosov flux lattice unstable, and one
therefore expects a first--order melting transition leading to a 
normal--conducting {\it flux liquid} phase with well below the upper
critical field $H_{\rm c_2}(T)$, at least for ideally pure systems
\cite{nelseu}.
It is very difficult to pin such a vortex liquid consisting of
fluctuating flexible lines by just intrinsic point disorder 
(e.g., oxygen vacancies); yet asymptotically a truly superconducting
low--temperature ``vortex glass'' phase has been predicted
\cite{fifihu}.

Therefore, in addition to point defects, the influence of extended or
{\it correlated} disorder, promising stronger pinning effects, on
vortex transport properties has been considered. 
Experimentally, linear damage tracks have been produced in oxide
superconductors by irradiation with high--energy ions.  
These {\it columnar defects} serve as strong pinning centers for the
flux lines, thus significantly increasing the critical current and
shifting the irreversibility line upwards (for references, see
Ref.~\cite{review}, Sec.~IX~ B).

In a long--wavelength description in the spirit of the London
approximation, one may consider the following model free energy for
$N$ flux lines, defined by their trajectories ${\vec r}_i(z)$ as they
traverse the sample, interacting with each other and with $N_{\rm D}$
columnar defects which are aligned along the magnetic--field direction
${\vec {\hat z}}$ \cite{drnvmv,nelvin}:
\begin{equation}
	F_N = \int_0^L dz \sum_{i=1}^N \Biggl\{ 
		{{\tilde \epsilon}_1 \over 2} 
		\left( {\D {\vec r}_i(z) \over \D z} \right)^2 +
    		{1 \over 2} \sum_{j \not= i}^N V[r_{ij}(z)] + 
		\sum_{k=1}^{N_{\rm D}} 
		V_{\rm D}[{\vec r}_i(z) - {\vec R}_k] \Biggr\} \ . 
\label{modelf}
\end{equation}
Here $r_{ij}(z) = | {\vec r}_i(z) - {\vec r}_j(z) |$, and 
$V(r) = 2 \epsilon_0 K_0(r / \lambda)$ denotes the repulsive vortex
interaction; the modified Bessel function ${\rm K}_0(x)\propto -\ln x$ 
for $x \to 0$, and $\propto x^{-1/2} \E^{-x}$ for $x \to \infty$. 
Thus the London penetration depth $\lambda$ defines the interaction
range and the energy scale $\epsilon_0 = (\phi_0 / 4 \pi \lambda)^2$.
For low fields $B \leq \phi_0 / \lambda^2$, i.e.: 
$\lambda \loa a_0 = (4/3)^{1/4}(\phi_0/B)^{1/2}$, the vortex line
tension is ${\tilde \epsilon}_1 \approx \epsilon_0$ ($a_0$ is the
triangular lattice constant in the corresponding pure system).
Finally, the pinning energy is a sum of $N_{\rm D}$ $z$--independent
potential wells $V_{\rm D}({\vec r})$ with average spacing $d$
centered on the positions $\{ {\vec R}_k \}$, whose diameters are
typically $b_0 \approx 100 \, \AA$, with a variation of 
$\delta b_k / b_0 \approx 15 \%$, caused by the ion--beam
dispersion. This induces some distribution $P$ of the pinning energies
$U_k$, which may be determined from the formula 
$U_k \approx (\epsilon_0 / 2) \ln [1 + (b_k / \sqrt{2} \xi)^2]$, where
$\xi$ denotes the coherence length \cite{nelvin}.
The thermodynamic properties of the model (\ref{modelf}) are to be
obtained by evaluation of the grand--canonical partition function
\begin{equation}
	{\cal Z}_{\rm gr} = \sum_{N = 0}^\infty {1 \over N !} \,
	\E^{\mu N L / T} \int \prod_{i = 1}^N {\cal D}[{\vec r}_i(z)] 
		\, \E^{- F_N[\{ {\vec r}_i(z) \}] / k_{\rm B} T} \ ,
\label{grpart}
\end{equation}
with the chemical potential $\mu = (H_{c_1} - H) \phi_0 / 4 \pi$,
which changes sign at the lower critical field 
$H_{c_1} = \phi_0 \ln (\lambda / \xi) / 4 \pi \lambda^2$.

As for pure systems \cite{nelseu}, one can map this statistical
problem of directed flux lines interacting with columnar pins, onto
the quantum mechanics of bosons in two dimensions subject to static
point disorder via the identification of the vortex trajectories
${\vec r}_i(z)$ with the particle world lines in imaginary time
\cite{drnvmv,nelvin,lyuksy}.
In Table~\ref{bosmap}, the corresponding quantities for bosons and
magnetic flux lines are listed (notice that the thermodynamic limit 
$L \to \infty$ for the directed polymers corresponds to a 
zero--``temperature'' quantum problem), and the ensuing phase diagram
is sketched schematically in Fig.~\ref{phadia}. 
At high temperatures, one thus finds an {\it entangled liquid} of
unbound flux lines (corresponding to a boson {\it superfluid}),
separated by a sharp second--order transition (see Ref.~\cite{fiswei}
and references therein) from a low--temperature phase of localized
vortices.
This {\it Bose glass} phase is characterized by an infinite tilt
modulus $c_{44}$, and turns out to be stable over a certain range of
tipping angles of ${\vec B}$ away from the $z$ direction
\cite{nelvin,davcon}.
At least for $\lambda \ll d$, the theory also suggests a
low--temperature {\it Mott insulator} phase, when $B = B_\phi$, i.e.,
the vortex density exactly matches that of the columnar pins. The Bose
glass and the Mott insulator are distinct thermodynamic phases, for
the latter should be characterized by infinite tilt {\it and}
compressional moduli \cite{nelvin}.

\begin{table}[htb]
\caption[]{Boson analogy applied to vortex transport.}
\begin{flushleft}
\renewcommand{\arraystretch}{1.2}
\begin{tabular}{lccccccccc}
\hline \noalign{\smallskip}
Bosons & \, $m$ & \, $\hbar$ & \, $\beta \hbar$ & \, $V(r)$ & \, $n$ &
	\, $\mu$ & \, $\rho_{\rm s}$ & \, ${\bf E}$ & \, ${\bf j}$ \\ 	
\noalign{\smallskip}
Vortices \, & ${\tilde \epsilon}_1$ & \, $T$ & \, $L$ & 
	\, $2 \epsilon_0 K_0(r / \lambda)$ & \, $B / \phi_0$ & 
	\, $(H - H_{\rm c_1}) \phi_0 / 4 \pi$ & 
	\, $\rho^2 c_{44}^{-1}$ & \, 
	${\bf {\hat z}} \times {\bf J} / c$ & \, ${\bf {\cal E}}$ \\
\noalign{\smallskip} \hline
\end{tabular}
\renewcommand{\arraystretch}{1}
\label{bosmap}
\end{flushleft}
\end{table}
\begin{figure}
\vspace{5.2 cm}
\caption[]{Schematic phase diagram for vortices interacting with
	columnar defects aligned along the magnetic--field direction.} 
\label{phadia}
\end{figure}

The formal analogy with a boson superfluid can be further exploited to
investigate the effect of disorder on density and current correlation
functions in the flux liquid phase within the Bogoliubov approximation 
\cite{nelled}. 
Thus one may compute the corresponding transport coefficients, e.g.,
the boson superfluid density $\rho_{\rm s}$ whose flux line analog is
the inverse tilt modulus of the vortex array (see Table~\ref{bosmap}). 
For parallel columnar defects of strength $U_0$ one eventually finds
\cite{tauber}:
\begin{equation}
	c_{44}^{-1}(T) \approx (n {\tilde \epsilon}_1)^{-1} 
		\Bigl[ 1 - \left( T_{\rm BG} / T \right)^4 \Bigr] ,
		\ {\rm with} \
	T_{\rm BG}(B) \approx T^* [\phi_0 / (4 \pi \lambda)^2 B]^{1/4}
\label{tilmod}
\end{equation}
and $k_{\rm B} T^* = b ({\tilde \epsilon}_1 U_0)^{1/2}$, in accord
with estimates obtained in the Bose glass itself \cite{nelvin}.
The critical behavior at the boson localization transition is not yet
fully understood, although quantum Monte Carlo simulations
\cite{walsor} seem to support the scaling theory \cite{fiswei}.
Direct simulations for the critical dynamics of flux lines near the
Bose glass transition have also been performed \cite{walgir}.

\section{Structural Properties in the Bose Glass Phase}

Recently, the positions of flux lines and parallel columnar pins,
aligned along the magnetic--field direction, were simultaneously
measured in a BSCCO sample in the Bose glass phase at low magnetic
fields using a combined chemical etching and magnetic decoration
technique \cite{dailie}.
Figure~\ref{cdflpo} depicts the thus obtained positions of $N_{\rm D}$
columnar defects and $N$ flux lines in a two--dimensional cross section 
perpendicular to ${\vec B}$ in a case where $f = N / N_{\rm D} = 
B / B_\phi \approx 0.24$ and $\lambda \approx 0.45 \, a_0$.
While the columnar defects are to a good approximation randomly 
distributed in space, the flux lines clearly form a highly 
{\it correlated} configuration; correspondingly, the two--dimensional
vortex structure function $S(q)$ displays a distinct peak at wave
vector $q \approx 2 \pi / a_0$, resembling an amorphous material
(see Fig.~\ref{strfac}).
This correlated structure must obviously be the result of the mutual
repulsive interactions $V_{\rm int}$ between the vortices which
dominate over any statistical fluctuations in the pinning potentials
$\delta V_{\rm D}$; on the other hand, all the flux lines are attached
to a defect, and therefore $V_{\rm D} > V_{\rm int}$ \cite{taudai}.

\begin{figure}
\vspace{6 cm}
\caption[]{Positions of empty columnar defects (open circles), and
	pins occupied by flux lines (filled circles), as obtained from
	a decoration experiment ($f \approx 0.24$).} 
\label{cdflpo}
\end{figure}

In order to model this experimental situation, we can fortunately
simplify the free energy (\ref{modelf}) substantially:
(i) As long as we limit ourselves to ``low'' temperatures (an estimate
for BSCCO actually yields $T \loa T_1= 0.9 \, T_c$) we can neglect
thermal wandering of the flux lines, which is furthermore
significantly suppressed by the confining line defects, we may drop
the tilt--energy $\propto {\tilde \epsilon}_1$ and consider an
effectively ``classical'' ($k_{\rm B} T \sim \hbar$ in the boson
analogy) two--dimensional system.
(ii) For magnetic fields smaller than the crossover field $B^*(T)$ in
Fig.~\ref{phadia}, the vortices are sufficiently dilute to ensure that
their repulsive interactions have no influence on the localization
length, and in fact the ``boson'' statistics becomes irrelevant (for
BSCCO, $B^*(T_1) \approx 0.7 \, B_\phi$).
Having thus disposed of the entropic corrections, we may consider the
following effective Hamiltonian and its grand--canonical counterpart
\cite{taudai,taunel}, 
\begin{equation}
	H = {1 \over 2} \sum_{i \not= j}^{N_{\rm D}} n_i n_j V(r_{ij})
		+ \sum_{i=1}^{N_{\rm D}} n_i t_i \ , \quad
	{\tilde H} = H - \mu \sum_{i=1}^{N_{\rm D}} n_i \ ,
\label{effham}
\end{equation}
where $i = 1,\ldots,N_{\rm D}$ denote the defect sites, randomly
distributed in the $x$-$y$ plane, $n_i = 0,1$ is the corresponding
site occupation number, and the $t_i$ are random--site energies
(originating in the varying pin diameters), whose distribution of mean
zero and width $w \approx 0.1 \l \epsilon_0$ we for simplicity choose
to be flat: $P(t_i) = \Theta \left( w - | t_i | \right) / 2 w$. 

This classical model is still far from trivial, however, due to the
interplay of long--range interactions and disorder.
Fortunately models of the form (\ref{effham}), albeit with a Coulomb
potential replacing the screened logarithmic interaction $V(r)$, have
been extensively studied in the context of disordered semiconductors,
which allows us to adapt qualitative arguments and numerical
simulation procedures from the literature (see, e.g., 
Refs.~\cite{shkefr,davlee}).
Thus, in order to infer the (approximate) ground--state properties in
the Bose glass phase as functions of the filling $f$ and interaction
range $\lambda / d$ we employ a zero--temperature Monte Carlo
algorithm that essentially minimizes the total energy (\ref{effham})
with respect to all possible one--vortex transfers; i.e., the ground
state stability is checked by demanding that all single--particle hops
from occupied sites $i$ to empty sites $j$ {\it require} an energy
$\Delta_{i \to j} = \epsilon_j - \epsilon_i - V(r_{ij}) > 0$, where
$\epsilon_i = t_i + \sum_{j \not= i} n_j V(r_{ij})$ are the 
{\it interacting} site energies.
For $\lambda \goa d$ and not too large $f \loa 0.6$ the resulting
distribution of flux lines among the columnar pins closely resembles
the experimental data in Fig.~\ref{cdflpo}. 
More quantitatively, we can investigate the amorphous peak at 
$q a_0 \approx 2 \pi$ in the vortex structure factor $S({\vec q}) =
\sum_{i,j} \E^{\I {\vec q} \cdot ({\vec r}_i - {\vec r}_j)} / N$,
averaged over all directions in reciprocal space. 
Whereas the dependence on the interaction range is rather weak as long
as $\lambda \goa d$, the peak in Fig.~\ref{strfac} vanishes when 
$f \goa 0.4$; this reflects that at higher fillings the system has to
increasingly accomodate with the underlying randomness, and therefore
the correlations induced by the interactions disappear.
Furthermore, a triangulation of the vortex positions shows that
typically only about 50 \% of the sites are sixfold coordinated, and
thus the structure formed by the flux lines rather resembles an
amorphous state (justifying the name ``Bose glass'') rather than a
weakly disordered triangular lattice \cite{taunel}.

\begin{figure}
\vspace{5.2 cm}
\caption[]{Dependence of the vortex structure factor peak on the
	filling $f$ for $\lambda / d = 2$; solid: $f = 0.1$,
	dot--dashed: $f = 0.2$, dashed: $f = 0.4$, and long--dashed:
	$f = 0.8$.}
\label{strfac}
\end{figure}

\section{The Distribution of Pinning Energies}

The above Monte Carlo algorithm can also be employed to determine the
distribution of pinning energies $g(\epsilon)$, with the vortex
interactions taken into account.
As in the boson picture this is nothing but the single--particle
density of states, $g(\epsilon)$ is readily obtained by sampling the
site energies $\epsilon_i$ for a number of disorder realizations and
then performing a quenched average.
Following the mean--field arguments of Efros and Shklovskii (1984) for
a system with long--range repulsive interactions 
$V(r) \propto r^{-\sigma}$, with $0 < \sigma < D$ (here, the space
dimension is $D = 2$), one expects that the density of states vanishes
at the chemical potential according to a power law,
\begin{equation}
	g(\epsilon) \propto | \epsilon - \mu |^{s_{\rm eff}} \ ,
\label{gapexp}
\end{equation}
with a {\it gap exponent} $s_{\rm MF} = (D / \sigma) - 1$.
This means that as a consequence of the correlations induced by the
long--range interactions, the probability of finding a low--energy
empty site ($\epsilon \goa \mu$) for introducing an additional
particle is greatly reduced, as is the density of states of
high--energy filled sites ($\epsilon \loa \mu$) for the corresponding
process of removing a particle from the system.
While the qualitative prediction (\ref{gapexp}) remains correct, the
actual gap exponent is influenced by subtle fluctuation effects
\cite{davlee} and therefore differs from the mean--field predictions
(for the currently most accurate results for Coulomb interactions,
$\sigma = 1$, see Ref.~\cite{mobius}).

\begin{figure}
\vspace{5.2 cm}
\caption[]{Normalized distribution of pinning energies 
	$G(E) = 2 \epsilon_0 d^2 g(\epsilon)$ as function of the 
	single--particle energies $E = \epsilon / 2 \epsilon_0$,
	with (a) $\lambda \rightarrow \infty$, and $f = 0.2$, 
	(b) $\lambda / d = 2$, and $f = 0.4$. 
	The location of $\mu$ is marked by the arrow.}  
\label{dosgap}
\end{figure}

Figure~\ref{dosgap} depicts two characteristic examples for the
distribution of pinning energies $g(\epsilon)$, as obtained from our
simulations \cite{taunel}, and normalized according to 
$\int g(\epsilon) \D \epsilon = 1 / d^2$.
In the case of an infinite--range logarithmic potential, the Ewald
summation method was used, and before taking the limit $\lambda \to
\infty$, a term  $\epsilon_i' - \epsilon_i = - 2 \pi f (\lambda / d)$
was subtracted from the site energies.
The filled and empty states are separated by a wide soft gap centered
at the chemical potential, with an effective gap exponent 
$s_{\rm eff} \approx 3$ for low filling.
Remarkably, this characteristic double--peak structure and the marked
depletion of states persists even for finite--range repulsion with
$\lambda \geq d$ (Fig.~\ref{dosgap}b), and although $g(\mu)$ does not
strictly vanish any more, the distribution of pinning energies for
$\epsilon \approx \mu$ can be fitted with a power law (\ref{gapexp});
e.g., $s_{\rm eff} \approx 1.2$ for $\lambda / d = 1$ and $f = 0.1$.
Upon increasing $f$, the gap closes quickly due to the stronger
influence of the underlying randomness (compare Fig.~\ref{strfac}),
but for $f \loa 0.2$ the correlation effects only disappear when
$\lambda < d$.

\section{Variable--Range Hopping Transport of Flux Lines} 

An in--plane current ${\vec J} \perp {\vec B}$ induces a Lorentz force
${\vec f}_{\rm L} = \phi_0 {\hat {\vec z}} \times {\vec J} / c$,
acting on the vortices. 
Accordingly, a term $\delta F_N = 
- \int_0^L dz \sum_i {\vec f}_{\rm L} \cdot {\vec r}_i(z)$ has to be 
added to the free energy (\ref{modelf}). 
In the spirit of the thermally assisted flux--flow (TAFF) model of
vortex transport, we write the superconducting {\it resistivity}
(i.e.: the {\it conductivity} in the boson representation) as $\rho = 
{\cal E} / J \approx \rho_0 \exp \left[ - U_{\rm B}(J) / T \right]$, 
where $\rho_0$ is a characteristic flux--flow resistivity, and 
$U_{\rm B}$ represents an effective barrier height. 
We shall see that in the Bose glass phase $U_{\rm B}(J)$ actually
diverges as $J \rightarrow 0$, 
\begin{equation}
	{\cal E} \approx \rho_0 J \exp 
	\left[ - E_0 / k_{\rm B} T) (J_0 / J)^{p_{\rm eff}} \right] \ . 
\label{curvol}
\end{equation}
\begin{figure}
\vspace{5.5 cm}
\caption[]{Saddle--point configurations for vortex transport in the
	Bose glass phase. (a) Vortex half--loop. 
	(b) Double--superkink in the variable--range hopping regime.}
\label{sadcon}
\end{figure}

Driven by an intermediate external current $J_1 < J < J_c$, a flux
line will start to leave its columnar pin by detaching a segment of
length $z$ into the defect--free region, thereby forming a half--loop
of transverse size $r$ (Fig.~\ref{sadcon}a). 
The free energy corresponding to this saddle--point configuration is
readily estimated to be $\delta F_1(r,z) \approx 
{\tilde \epsilon}_1 r^2 / z + U_0 z - f_{\rm L} r z$, and optimizing 
successively with respect to $z$ and $r$ one finds $z^* \approx 
r^* \sqrt{{\tilde \epsilon}_1 / U_0} \approx 
c \sqrt{{\tilde \epsilon}_1 U_0} / \phi_0 J$. 
Hence the current--voltage characteristics assumes the form
(\ref{curvol}), with 
$E_0 = E_{\rm K} = d \sqrt{{\tilde \epsilon}_1 U_0}$, 
$J_0 = J_1 = c U_0 / \phi_0 d$, and the effective 
{\it transport exponent} $p_{\rm eff} = 1$ in the 
{\it vortex half--loop regime} \cite{nelvin}. 
For $J \to J_1$, the flux line will extend a double kink of width
$w_{\rm K} = d \sqrt{{\tilde \epsilon}_1 / U_0}$ and energy 
$2 E_{\rm K}$ to an adjacent columnar defect, and therefore for 
$J < J_1$ neighboring pins must be taken into account.

Thus, at low currents one enters another regime, where the flux line
emits a pair of kinks to a possibly distant defect which happens to
provide a favorable pinning energy (Fig.~\ref{sadcon}b).
This is the vortex analog of variable--range charge transport in
disordered semiconductors (see Ref.~\cite{shkefr} and references
therein).
The cost in free energy for such a configuration of transverse size
$R$ and extension $Z$ along the magnetic--field direction will consist
of the double--kink energy $2 E_{\rm K}(R) = 2 E_{\rm K}(d) R / d$
stemming from the elastic term, and the difference in pinning energies
of the highest--energy occupied site, $\epsilon_i \approx \mu$, and
the empty site at distance $R$ with $\epsilon_j = \mu + \Delta(R)$,
where the distribution of pinning energies enters through the condition
$\int_\mu^{\mu + \Delta(R)} g(\epsilon) \D \epsilon = R^{-D}$.
Minimizing the saddle--point free energy
$\delta F_{\rm SK} \approx 2 E_{\rm K} R / d + Z \Delta(R) - f_L R Z$
in the regime where $g(\epsilon)$ has the form (\ref{gapexp}), one
arrives at (\ref{curvol}) with $E_0 = 2 E_{\rm K}$ and the effective
transport exponent \cite{taunel}
\begin{equation}
	p_{\rm eff} = (s_{\rm eff} + 1) / (D + s_{\rm eff} + 1) \ .
\label{ivexpo}
\end{equation}
\begin{figure}
\vspace{5.2 cm}
\caption[]{Log--log plots (base 10) of the activation energy 
	$U_{\rm B}(J) \propto R^*(J) / d$ for variable--range hopping
	vs the reduced current $j = J \phi_0 d / 2 \epsilon_0 c$ for
	$f = 0.1$. 
	Diamonds: $\lambda \rightarrow 0$, circles: $\lambda / d = 1$,
	squares: $\lambda / d = 2$, triangles: $\lambda / d = 5$.}
\label{ivchar}
\end{figure}

For short--range interactions, $g(\mu) > 0$, i.e., $s_{\rm eff} = 0$,
and the Mott variable--range hopping exponent 
$p_{\rm M} = 1 / (D + 1)$ is recovered \cite{nelvin}.
But as the interaction range $\lambda$ becomes larger than the average
vortex separation $a_0 = d / \sqrt{f}$, the emerging correlations
reduce the phase space for single--vortex hopping drastically, and
$p_{\rm eff} < p_{\rm M}$; in addition, as the normalized distribution
of pinning energies is broadened by the interactions, also the current
scale $J_0$ in (\ref{curvol}) is increased.
The result of the minimization procedure using the numerical data for
$g(\epsilon)$ in the case $f = 0.1$ is shown in Fig.~\ref{ivchar} as a
function of the interaction range \cite{taunel}.
Assuming that flux line transport in the low--current regime is still
dominated by single--vortex hops, this implies that long--range
repulsive interactions reduce the resistivity by several orders of
magnitudes with respect to the non--interacting situation.
E.g., for $\log j \approx 1.5$ the exponent in (\ref{curvol}) is about
ten times smaller for $\lambda / d = 5$ as compared to 
$\lambda \rightarrow 0$ (see Fig.~\ref{ivchar}), and hence $\rho$ is
reduced by a factor of $\approx 10^{-5}$.
In accord with the relation (\ref{ivexpo}), the effective transport
exponent assumes values upto $p_{\rm eff} \approx 0.7$ for 
$\lambda \to \infty$ and $f = 0.1$, while, e.g., for $\lambda / d = 1$
and $f = 0.1$ we find $p_{\rm eff} \approx 0.55$.
We remark that the former value for for the variable--range hopping
exponent of logarithmically interacting particles in two dimensions is
consistent with an analysis based on the very different assertion that
actually {\it collective} many--particle hops yield the lowest energy
barriers \cite{fistok}.
Finally, a recent experiment on a BSCCO sample with $f \approx 0.15$
and $\lambda / d \approx 1.6$ reported a measured 
$p_{\rm eff} \approx 0.57$ in the variable--range hopping regime 
across about half a decade in the current--voltage characteristics
\cite{konzyk}; this apparently agrees well with our results, although
due caution needs to be applied for this comparison as in this
experiment the magnetic field distribution was in fact very
inhomogeneous.
Yet further experiments should clearly be capable to test the above
predictions quantitatively.

In summary, experiments have established that columnar defects
provide very effective pinning centers for flux lines in high--$T_c$
superconductors, and demonstrated that in certain cases the repulsive
vortex interactions lead to remarkable spatial correlations.
As a consequence, a soft ``Coulomb gap'' emerges in the distribution
of pinning energies, and vortex transport in the variable--range
hopping regime should be drastically reduced.

\subsubsection{Acknowledgments.}
The work presented here was done in delightful collaboration with
David Nelson, whose ideas and insights have been invaluable. 
I would also like to thank H.~Dai and C.M.~Lieber for fruitful
cooperation and their sharing of data with us prior to publication. 
This research was supported by the Deutsche Forschungsgemeinschaft
under Contract Ta.~177/1, and by the National Science Foundation, in
part by the MRSEC Program through Grant DMR-9400396, and through Grant
DMR-9417047.
Financial support from a Royal Society Conference Grant and from the
Engineering and Physical Sciences Research Council through Grant
GR/J78327 is gratefully acknowledged.


\begin{thebibliography}
\bibitem{[8]}{davcon}{[8]}
	Further details on this ``transverse Meissner effect'' and
	vortex localization in transverse magnetic fields may be found
	in the contribution to this volume by D.R.~Nelson.
\end{thebibliography}
\end{document}